\documentclass[prl,twocolumn,showpacs]{revtex4}
\usepackage[tbtags]{amsmath}
\usepackage{graphicx}
\usepackage{bm}

\def\beq{\begin{equation}}
\def\eeq{\end{equation}}
\def\eeql#1{\label{#1} \end{equation}}
\def\C{\Gamma}
\def\D{\Delta}
\def\W{\Omega}
\def\c{\gamma}
\def\e{\epsilon}
\def\w{\omega}

\def\im{\mathop{\rm Im}\nolimits}
\def\re{\mathop{\rm Re}\nolimits}
\def\Ei{\mathop{\rm Ei}\nolimits}
\newcommand{\To}{\rightarrow}

\begin{document}

\title{Unconventional Gravitational Excitation of a Schwarzschild Black Hole}

\author{P.T. Leung}
\author{Alec \surname{Maassen van den Brink}}
\thanks{Corresponding author;\\ electronic address: \texttt{alec@dwavesys.com}}
\author{K.W. Mak}
\author{K.~Young}
\affiliation{Physics Department, The Chinese University of Hong Kong, Hong Kong, China}

\date{\today}


\begin{abstract}
Besides the well-known quasinormal modes, the gravitational spectrum of a Schwarzschild black hole also has a continuum part on the negative imaginary frequency axis. The latter is studied numerically for quadrupole waves. The results show unexpected striking behavior near the algebraically special frequency $\W=-4i$. This reveals a pair of unconventional damped modes very near~$\W$, confirmed analytically.
\end{abstract}

\pacs{04.30.-w
, 04.70.Bw
}

\maketitle


Black-hole linearized gravitational waves propagate on the spacetime around the event horizon. Because waves escape to infinity and into the event horizon, the
system is dissipative, and described by its nontrivial spectrum in the lower-half frequency plane. The latter consists of well-studied quasinormal modes (QNMs) \cite{Chandra,Ferrari,Nollert,Liu}, and a cut along the negative imaginary axis (NIA) giving rise to a continuum, about which little is hitherto known.

The Schwarzschild black hole is the simplest compact object in relativity, giving its gravitational excitation spectrum (part of which should soon be observable~\cite{LIGO}) a fundamental status akin to the hydrogen problem in quantum mechanics. Interest in this (classical) spectrum is surging~\cite{hod}, as it seems to offer clues to the quantum theory, in particular to a calculation of the Bekenstein entropy in loop quantum gravity and to the quantum of area. A correspondence between classical frequencies and quantum excitations should not be surprising---cf.\ the harmonic oscillator. We shall be content to have these developments serve as context and motivation.

In this article, the continuum spectrum is evaluated and characterized. It is found to oscillate in $\im\w$ (see Fig.~\ref{fig2} below), with a half-period equal to the spacing~\cite{asy} of the string of QNMs parallel and close to the NIA that are studied in Ref.~\cite{hod}. In particular, for the principal case of waves propagating from a source near the horizon to a distant observer, the spectrum is sharply dominated by a dipole contribution, revealing a pair of QNMs $\w_\pm$ on the unphysical side of the cut, very close to the algebraically special frequency~$\W$~\cite{couch}. This represents \emph{unconventional} damped modes. Their effect is more than indirect: as the black hole is given some rotation, $\w_\pm$ seem to cross the NIA and emerge onto the physical sheet. This follows from precise and nontrivial numerical studies, combined with analytics at and near~$\W$.

For a black hole of mass $M$ ($c = G = 2M=\nobreak1$ below) and each angular momentum $\ell$, the radial functions $\psi$ of axial gravity waves are governed by 
a generalized Klein--Gordon or so-called Regge--Wheeler equation
\beq
  [d_x^2 + \w^2 - V(x)]\psi(x,\w)=0\;;
\eeql{RWE}
$x=r+\ln (r{-}1)$ is the tortoise coordinate and $r$ the circumferential radius. The potential $V(r)=(1{-}r^{-1})\*[\ell(\ell{+}1)r^{-2} - 3r^{-3}]$ accounts for the Schwarzschild background~\cite{R-W}. We impose outgoing-wave conditions (OWCs) $\psi(x{\To}{\pm}\infty,\w) \sim e^{i\w |x|}$. (For $x\to-\infty$, waves thus go into the horizon.)

Polar waves obey the Zerilli equation~\cite{Zerilli}---a Klein--Gordon equation with $\tilde{V}(x)=V(x)+2d_xW(x)$, with
\beq
  W(r) = \C + \frac{3(r-1)}{r^{2}[(\ell{-}1)(\ell{+}2)r+3]}\;,
\eeql{eq:suppot}
$\C=(\ell{+}2)!/[6(\ell{-}2)!]\equiv i\W$. Below, we take $\ell=2$, the most important case, so $\C=4$. The solutions $\psi$ of (\ref{RWE}) and $\tilde{\psi}$ of the Zerilli equation are related by ``intertwining'' or supersymmetry: $\tilde{\psi}(x,\w) = [ d_x + W(x) ] \psi(x,\w)$ \cite{Chandra,chand,Wong}. Thus, also the two continua are closely related, and we focus on~(\ref{RWE}).

At $\W$, (\ref{RWE}) has an exact solution $g(x,-\W)\propto\exp\{-\int^x\!dy\,W(y)\}$~\cite{chandra84}; cf.\ below (\ref{eq: green}) for $g(x,\w)$. The nature of $\W$ has sometimes been controversial, with both a QNM~\cite{leaver1} and---in apparent contradiction---a total-transmission mode~\cite{and} having been claimed to occur. Moreover, the numerical spectra \emph{near} $\W$ for small Kerr-hole rotation $a$ point to intricate behavior for $a\downarrow0$~\cite{ono}. The issues were resolved in Ref.~\cite{Alec}: (a)~\emph{At} $\W$, (\ref{RWE}) and the Zerilli equation cannot be identified, for the supersymmetry transform relating them is singular there. (b)~Even for each separate sector (axial or polar), the OWCs at $\W$ are highly subtle [cf.\ below~(\ref{eq: green})]; only with precise definitions can the mode situation be elucidated. (c)~For $a\ne0$, $\W$ splits into distinct QNM \emph{and} total-transmission-mode multiplets; still, questions remain, to which we will return in the Discussion. This paper vindicates and augments these findings by uncovering the pair of nearby unconventional QNMs~$\w_\pm$. Besides being of fundamental significance, this further explains the problem's potential for numerical artifacts unless special precautions are taken. First, some preliminaries must be presented.

\begin{figure}
  \includegraphics[height=3in,angle=270]{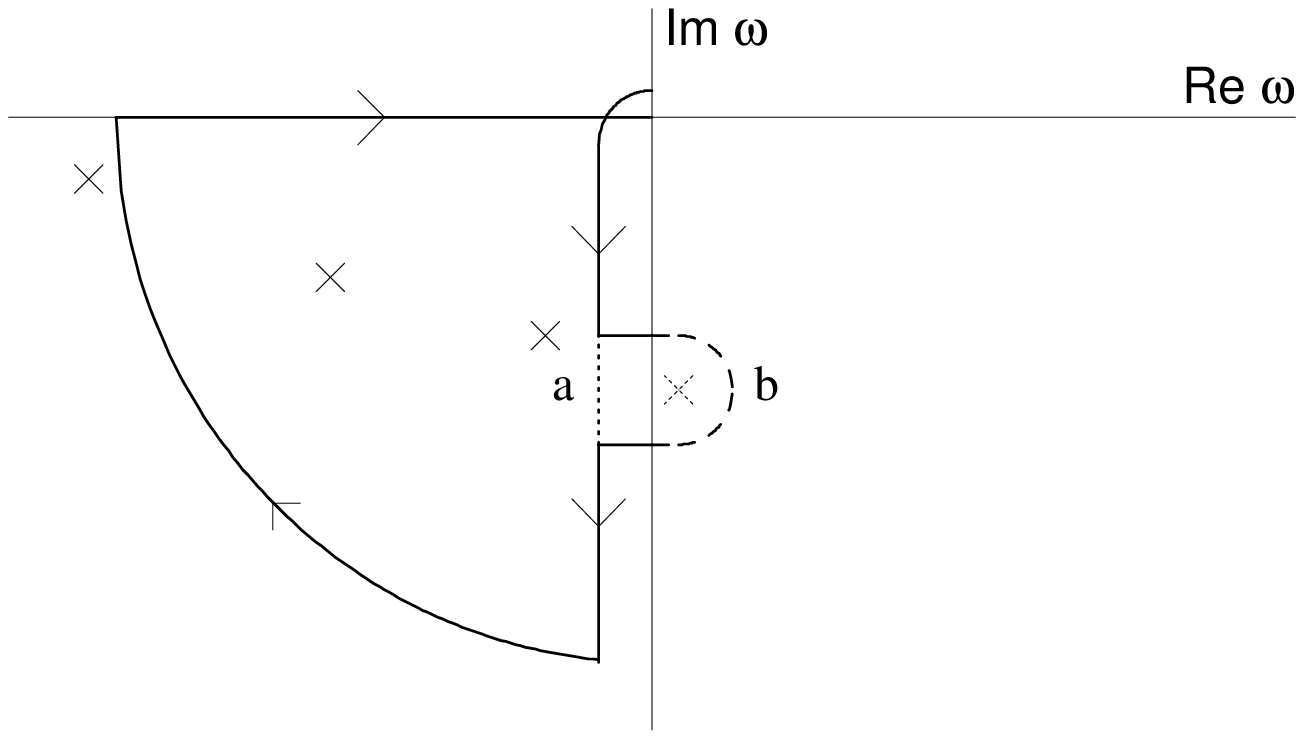}
  \caption{Fourier inversion of $\bar{G}$; crosses denote poles.
(a)~Conventional contour involving a cut on the NIA.
(b)~Modified contour detouring around the unconventional pole $\w_-$ near~$\W$ (contour and pole on unphysical sheet shown with broken lines). For simplicity the mirror contour is not shown.}
  \label{fig1}
\end{figure}

The Green's function $G(x,y;t) = \int (d\w /2\pi)\,e^{-i\w t}\*\bar{G}(x,y;\w)$ relates $\psi(x,t {\ge} 0)$ to $\{\psi(y,0),\dot{\psi}(y,0)\}$. Closing the contour in the lower-half $\w$\nobreakdash-plane (Fig.~\ref{fig1}, line~a) splits $\bar{G}$ into
(a)~the large semicircle, giving a prompt signal propagating directly from $y$ to $x$ and vanishing after a finite~$t$ \cite{Ching-cut,Bachelot};
(b)~the QNM poles, causing a ringing signal dominating at intermediate~$t$ \cite{Leaver,Ching-QNM}; and
(c)~our main focus, the branch cut on the NIA:
\beq
  \D \bar{G}(x,y;-i\c) =\bar{G}_{+}(x,y;-i\c)-\bar{G}_{-}(x,y;-i\c)\; ,
\eeql{eq:cut}
where $\bar{G}_\pm(-i\c)=\lim_{\e \downarrow 0} \bar{G}(-i\c {\pm}\e)$ are continuations from $\pm\w>0$. The physical sheet for $\bar{G}_{+}$ ($\bar{G}_{-}$) thus lies to the right (left) of the NIA.  The continuum is given by $\D \bar{G}$, and for $\c\downarrow0$ causes the late-$t$ behavior \cite{Leaver,Ching-cut}.

In general, the Green's function $\bar{G}(x,y)=\bar{G}(y,x)$ is
\beq
  \bar{G}(x,y;\w)=J^{-1}(\w)f(y,\w)g(x,\w)\; , \qquad y<x \; ,
\eeql{eq: green}
where $f$ ($g$) solves (\ref{RWE}) with the left (right) OWC, and $J=gf'-fg'$ is their Wronskian. We define $f(x{\rightarrow}{-}\infty,\w) \sim 1 \cdot e^{-i\w x}$ and $g(x{\rightarrow}\infty, \w) \sim 1 \cdot e^{i\w x}$. At a zero of~$J$, $f \propto g$ satisfies both OWCs and defines a QNM.

If $V$ has its support in say $[-d,d]$, the OWCs can be imposed at $\pm d$, so the wave equation is integrated over a finite length; hence, $f,g$ are analytic in $\w$. If however $V(x{\To}{-}\infty)\sim\sum_kv_ke^{\lambda kx}$ (here $\lambda=\nobreak1$), Born approximation shows that $f$ typically has poles (\emph{anomalous points}) at $\w_n=-in\lambda/2$. These are removable by rescaling $f(\w)\mapsto(\w{-}\w_n)f(\w)$, leaving $\bar{G}$ unaffected. However, $\{v_k\}$ could ``miraculously" leave some $f(\w_n)$ non-singular. For (\ref{RWE}), $n = \nobreak2\C$ is the only miraculous point (for any~$\ell$), which can be studied analytically~\cite{Alec}.

\begin{figure}
  \includegraphics[height=3in,angle=270]{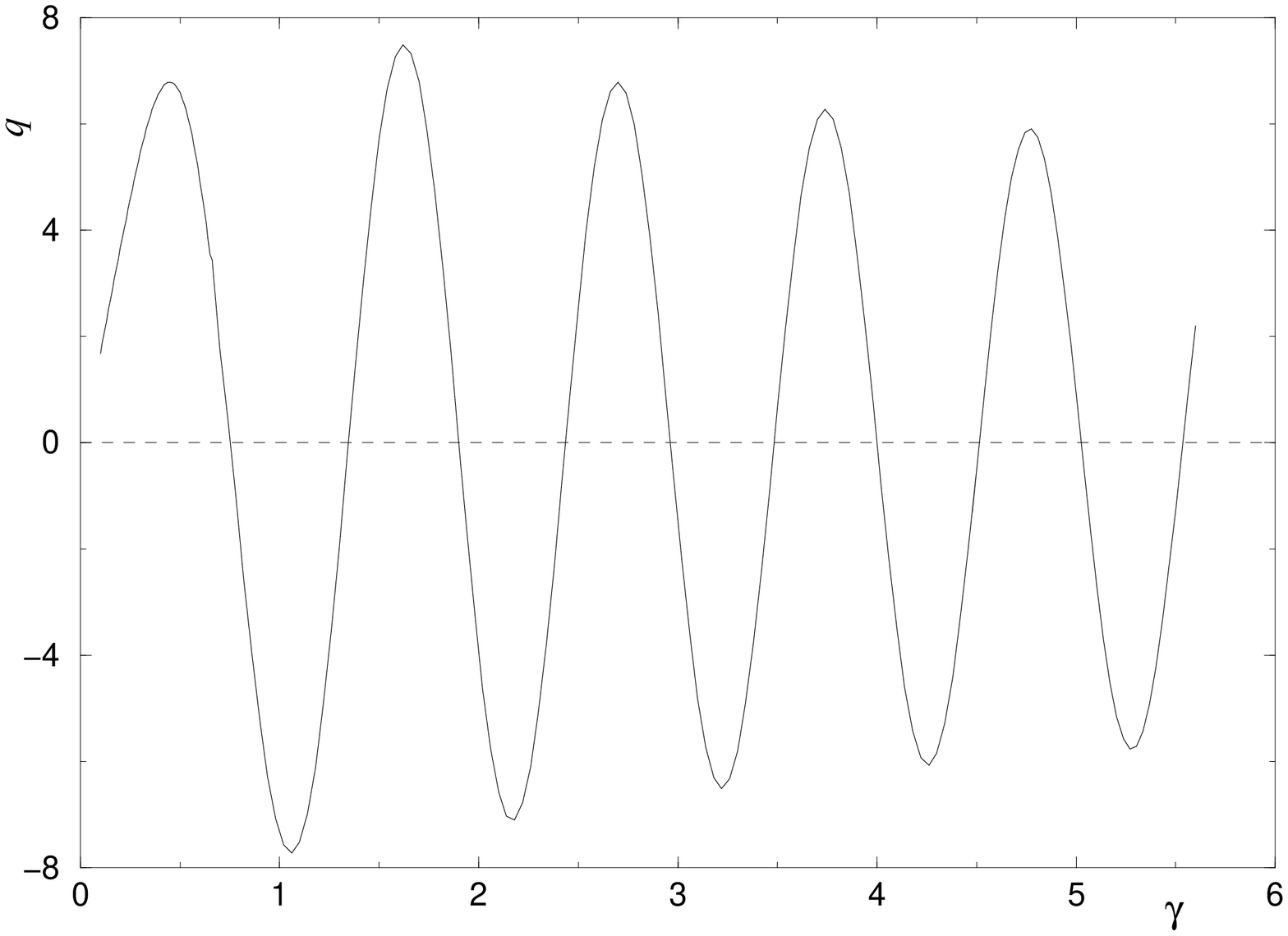}
  \caption{Plot of $q(\c)$.}
  \label{fig2}
\end{figure}

\paragraph*{Cut strength.}
For $x \To +\infty$, the centrifugal barrier does not scatter~\cite{Ching-cut}, so consider ($\c_\mathrm{E}$ is Euler's constant)
\beq
  V(x)-\frac{6}{x^2} \sim \frac{\ln x}{x^3}
  = \int_{0}^{\infty}\!\!\! d\lambda\, (3{-}2\c_\mathrm{E}{-}\ln\lambda)
  \lambda^{2}e^{-\lambda x}\;.
\eeql{Vtail}
The superposition of tails $e^{-\lambda x}$ spreads the poles $-in\lambda/2$ into a cut. On the NIA, $g_{\pm}\sim1 \cdot\nobreak e^{\c x}$ obey the same wave equation~(\ref{RWE}). Hence $\D g \sim 0 \cdot e^{\c x}$ [cf.~(\ref{eq:cut}) for~$\D$] is the small solution $\propto\nobreak g(+i\c)$. Since $g(-\w^*)=g^*(\w)$, one has $\re\D g=\nobreak0$. The cut strength $q(\c)\equiv\D g(x,-i\c)/[ig(x,+i\c)]$~\cite{Alec} is independent of $V(x)$ at any finite $x$: if, say, $V_1(x{>}L) = V_2(x{>}L)$, then $q_1=q_2$. Thus, $q$ efficiently characterizes the cut in $g$, causing the one in $\bar{G}$; cf.~(\ref{eq: DG and q-2}). Ref.~\cite{Leaver} has an expression for~$q$, but it is hard to evaluate.

We integrate $g(x,+i\c)$ from large $x$, but for $g(x,-i\c)$ this would be unstable~\cite{Tam}. Instead, we find $g(x,-i\c {\pm} \e)$ by Miller's algorithm of downward recursion on Leaver's series~\cite{long,Leaver series,YTLiu}, taking $\e \downarrow 0$ in the difference for $\D g$ (Miller's algorithm fails at $\e=\nobreak0$). Fig.~\ref{fig2} shows the ensuing $q(\c)$, consistent with $q'(0)=2\pi$~\cite{Ching-cut}. This and subsequent figures constitute an independent look at the Schwarzschild spectrum down in the complex $\w$-plane, now thought to have deep physical significance~\cite{hod}.

The zeros at $\c=$ 0.75, 1.35, 1.90, 2.44, 2.96, 3.48, 4.00, 4.51, 5.03, 5.54, \ldots\ suggest that the spacing approaches~$\frac{1}{2}$. Indeed $q(\c) \sim 4 \cos(2\pi\c) + O(\c^{-1/2})$ by a WKB analysis~\cite{Alec2}. The zero at $\C$ and moreover $q'(\C) \approx -37.6$ agree with~\cite{Alec}
\beq
  -q'(\C)=\frac{45\pi(2100027e^{8}+30148389005)}{137438953472}
  \approx 37.45\;,\label{eq:q-slope}
\eeq
confirming both (\ref{eq:q-slope}) and our numerical accuracy.  

\begin{figure}
  \includegraphics[height=3in,angle=270]{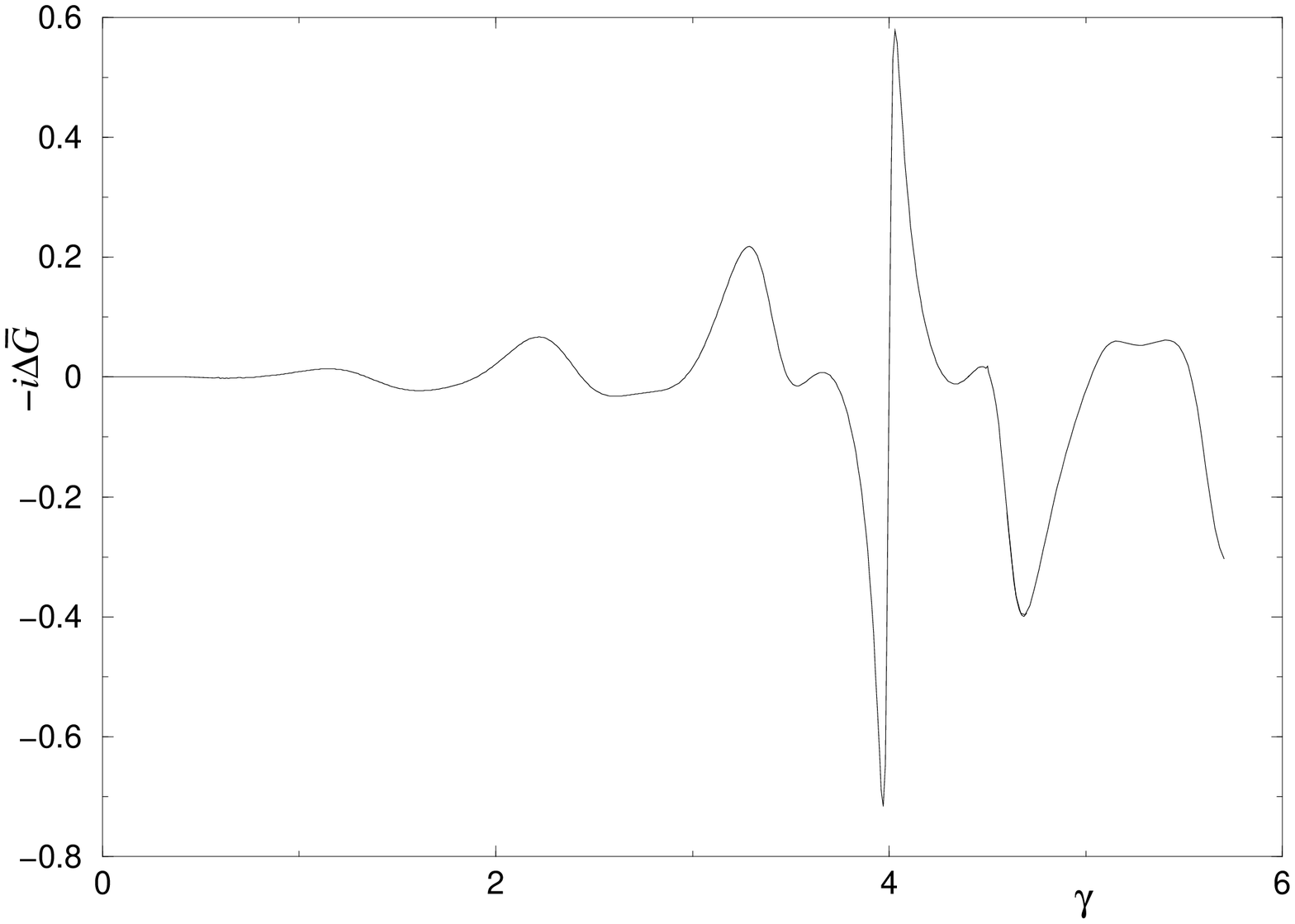}
  \caption{Plot of $-i \D \bar{G}(x,y;-i\c)$ for $x=0.2$, $y=0.1$.}
  \label{fig4}
\end{figure}

The cut in $\bar{G}$ itself is
\beq
  \D\bar{G}(x,y;-i\c)
  = -2i\c q(\c)\frac{f(x,-i\c)f(y,-i\c)}{J_+(-i\c)J_-(-i\c)}
\eeql{eq: DG and q-2}
[cf.\ below~(\ref{eq:cut}) for the $\pm$ subscripts]. Although $f$ can be stably integrated when matching to a Born approximation instead of to $f(z{\To}{-}\infty, -i\c) \sim e^{-\c z}$~\cite{Tam}, we prefer Jaff\'e's series \cite{Jaffe,Leaver series}. Figure~\ref{fig4} shows a typical result. For some refinements, see Ref.~\cite{long}. For $\c\downarrow0$, (\ref{eq: DG and q-2}) reproduces $G(x,y;t{\To}\infty)$~\cite{Ching-cut}.

\begin{figure}
  \includegraphics[height=3in,angle=270]{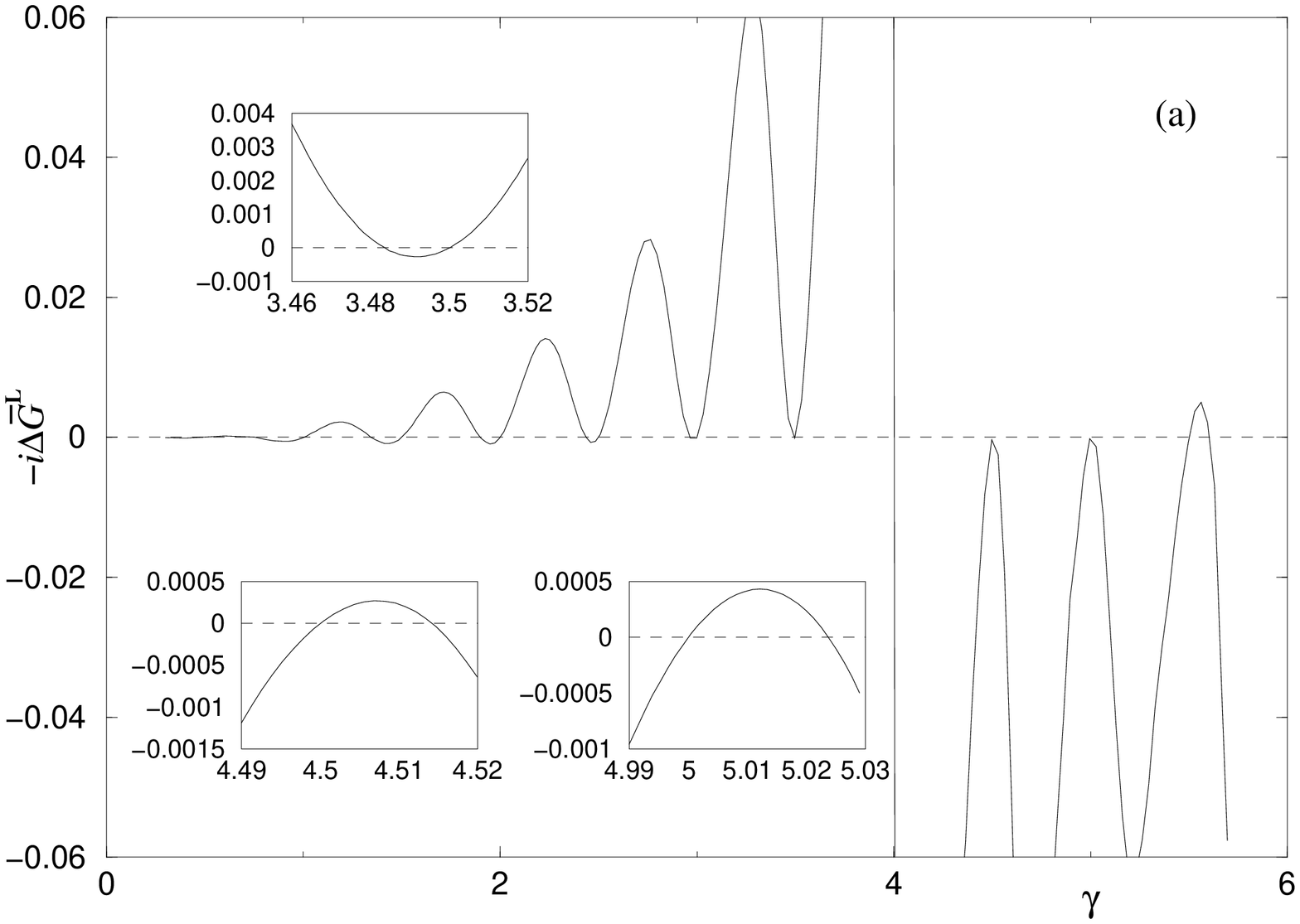}

  \vspace{1mm}
  \includegraphics[height=3in,angle=270]{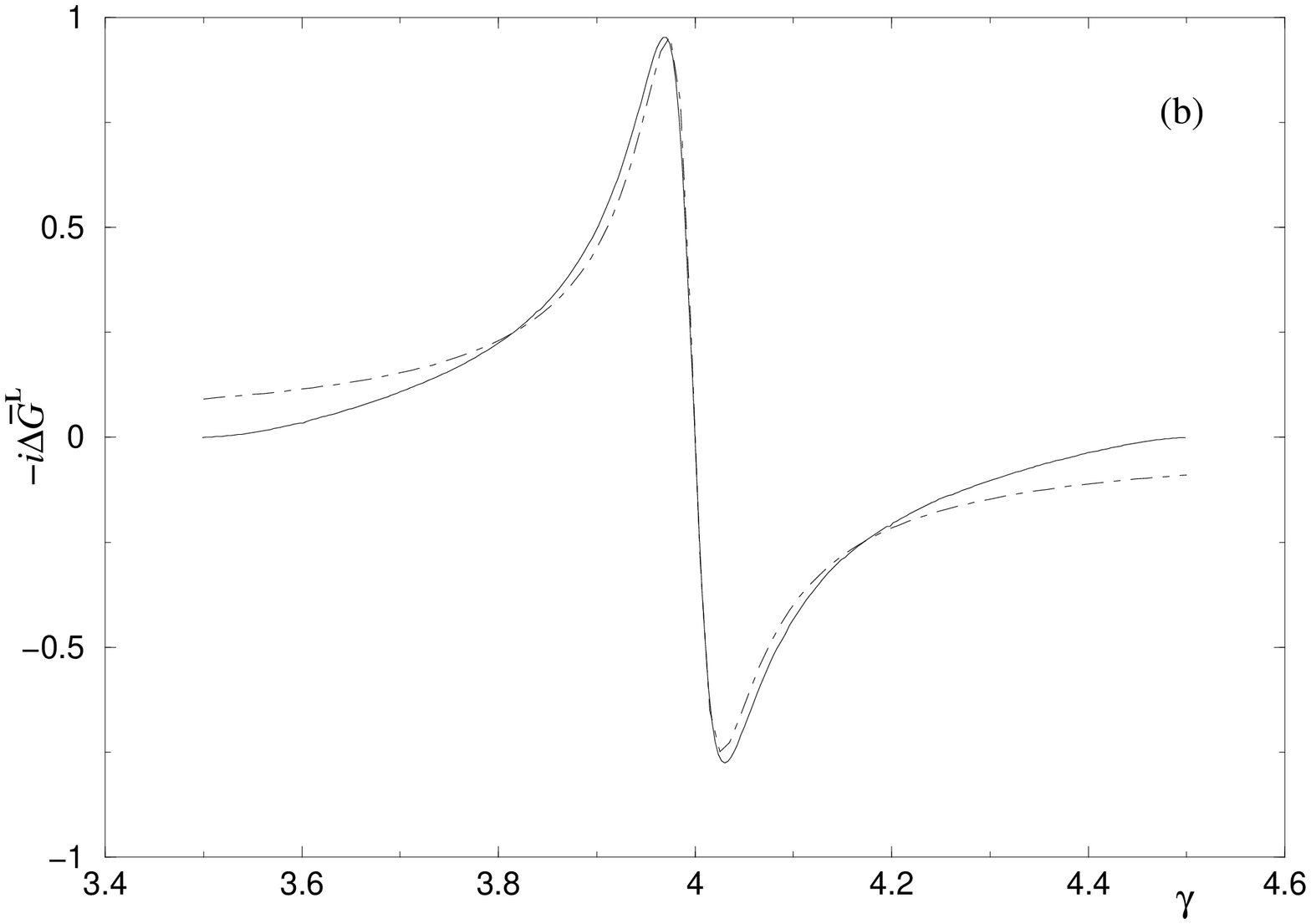}
  \caption{Plots of $-i \D \bar{G}^{\rm L}(-i\c)$.
(a)  For $0 \le \c \le 5.6$.  Insets show some regions with two close zeros.
(b)  The region $\c \approx 4$ expanded. The solid line is the numerical result, and the broken line is a fit to (\ref{eq:fit1}) with $a_2=-0.0227$.}
  \label{fig5}
\end{figure}

The important limit is $x,-y \rightarrow \infty$. Since $f(y, -i\c) \sim e^{\c|y|}$ and in general also $f(x, -i\c) \sim\text{const}\cdot e^{\c x}$, (\ref{eq: DG and q-2}) has a strong position dependence---simply a result of the long propagation time, and removed if $t$ is measured from the first arrival at $t_0(x,y) \equiv x-y$. Define $G^{\rm L}(t') \equiv \lim_{x, -y \rightarrow \infty} G(x,y; t_0 {+} t')$, so $\bar{G}^{\rm L}=J^{-1}$ by (\ref{eq: green}) and the normalization of $f,g$. Thus, $\D\bar{G}^{\rm L}=J_+^{-1}-J_-^{-1}$; see Fig.~\ref{fig5} for results.

The cut $\D \bar{G}^{\rm L} (-i\c)$ vanishes at (a) the zeros of $q(\c)$ [cf.~(\ref{eq: DG and q-2})], and (b)~the anomalous points $\c = \frac{1}{2}, 1, \frac{3}{2}, \ldots$, where $f$ and hence $J$ diverge.  The former depend \emph{only} on $V(x{\rightarrow} {+}\infty)$ [cf.\ below (\ref{Vtail})], the latter \emph{only} on $V(x{\rightarrow} {-}\infty)$, scaling with $\lambda$. If the two tails are separately adjusted the sequences are independent, but for $\lambda=1$ they share one member at~$\C$. Some members of the two sequences are close, cf.\ the insets in Fig.~\ref{fig5}a. The many zeros also keep $\D \bar{G}^{\rm L}$ small generically.

\paragraph*{Unconventional QNM.}
Although $\D \bar{G}^{\rm L}(-i\C)=0$, surprisingly $\D \bar{G}^{\rm L}$ is largest (close to a dipole) \emph{near} $\c = \C$ (Fig.~\ref{fig5}b; the dominance is less pronounced for finite $y$, cf.\ Fig.~\ref{fig4}). This reveals a pair of nearby QNM poles $\w_\pm$: if $\bar{G}^{\rm L}_{\pm}(\omega)\approx (a_1 {\pm}ia_2)/(i\omega{-}4{-}b{\pm}ic)$, with $a_1, a_2, b$ real and $c>0$ so $\w_\pm$ are on the unphysical sheets, then
\beq
  \D \bar{G}^{\rm L} (-i\c) \approx
  \frac{2ia_2(\c-\C)}{(\c{-}\C{-}b)^2 + c^2}\;.
\eeql{eq:fit1}
Here, $a_1 c + a_2 b = 0$ enforces the zero at $\c = \C$.  The broken line in Fig.~\ref{fig5}b shows this fit, yielding
\beq
 \w_\pm + i\C = \mp c - ib \approx  \mp 0.027 + 0.0033i\; .
\eeql{eq:position}

The extrapolation into the unphysical sheet can also be carried out analytically, by assuming that $J_+(\w{\approx}\W)$ can be linearized up to the nearest zero,
\beq
  \w_++i\C\approx-\frac{J(\W)}{J'_+(\W)}\;.
\eeql{om-J}
Since $\D J(\W)=0$, the sheet of~$J(\W)$ need not be indicated. Following the methods of~\cite{Alec}, one readily obtains $J(\W)=-\frac{700009}{917504}$. Further, a lengthy calculation gives~\cite{long}
\begin{align}
  J'_+(\W)=\frac{i}{49\cdot2^{38}}\bigl[&{-}17122265640585(\c_{\rm E}{+}\ln8{-}i\pi)\notag\\[-1mm]
    &-245810518235861775\Ei(8{+}i\eta)e^{-8}\notag\\
    &+36326230655979688\bigr]\;,\label{Jres}
\end{align}
where $\Ei(z)\equiv\int_{-\infty}^zdt\,e^t\!/t$ and $\eta>0$ is an infinitesimal. Insertion into (\ref{om-J}) yields
\beq
  \w_++i\C\approx-0.03248+0.003436i\;;
\eeql{om*full}
especially the agreement of $\im\w_+$ with (\ref{eq:position}) is remarkable. Since the latter value is also found by extrapolation, it need not be more accurate.

\paragraph*{Discussion.}
Like the hydrogen spectrum, the Schwarzschild spectrum contains discrete \emph{and} continuum parts. While the former is a classic of physics, the latter is much more difficult, being spheroidal rather than hypergeometric~\cite{Leaver series}. We have characterized the continuum, recovering the behavior both for $\gamma\downarrow0$ and near the miraculous point~$\C$. This leads to the identification of an essentially new type of damped excitation $\w_\pm$, which clearly affects the dynamics more than QNMs on the physical sheet at larger $\left|\im\w\right|$~\cite{hod}. An obvious question is whether there are further singularities on the unphysical sheets, and what is their influence on the cut.

In a wider context, consider the Kerr hole. By comparing numerics for moderately small $a$~\cite{ono} with the QNM multiplet found analytically to branch off from $\W$ at $a=0$, one concludes that one \emph{additional} multiplet has to emerge (as $a$ increases) near~$\W$. Rather than the possibilities considered in Ref.~\cite{Alec}, this multiplet may well be due to $\w_\pm$ splitting (since spherical symmetry is broken) and moving through the NIA as $a$ is tuned. As a first step, the continuum should be studied also for $a>0$.

One is led to consider another Fourier contour going into the unphysical sheet and detouring around~$\w_\pm$ (Fig.~\ref{fig1}, line~b), including them as QNMs. This slightly reduces the continuum (often neglected as ``background")~\cite{long}. More importantly, if these poles emerge onto the physical sheet when tuning a parameter (say, $a$), the total QNM and continuum contributions now become \emph{separately} continuous.

All of the above refers to $\ell = 2$. The much larger $\c\approx\C(\ell{\ge}3)$ are still unattainable numerically, but since any unconventional poles can be shown to be further away from $\W$~\cite{long}, their influence on the cut should be smaller.

These questions may be explored through solvable models with potential tails. Some aspects of the Regge--Wheeler equation can also be analyzed asymptotically~\cite{Alec2}. Numerical algorithms valid on the NIA and even into the unphysical region would also be useful, allowing QNMs there to be studied directly rather than through extrapolation.


\begin{acknowledgments}
We thank E.S.C. Ching, Y.T. Liu, W.M. Suen and C.W. Wong for many discussions, and the Hong Kong Research Grants Council for support (CUHK 4006/98P). AMB was also supported by a C.N. Yang Fellowship.
\end{acknowledgments}


\end{document}